\documentclass{aa}  

\usepackage{graphicx}
\usepackage[normalem]{ulem}
\usepackage{xcolor}
\usepackage{txfonts}
\usepackage{multirow}
\usepackage{array}
\usepackage{array}
\newcolumntype{C}[1]{>{\centering\arraybackslash}m{#1}}
\usepackage{stfloats}
\usepackage{subcaption}
\usepackage{lscape}
\usepackage{placeins}
\usepackage{natbib}
\bibpunct{(}{)}{;}{a}{}{,}                        

\begin{document}
   
   \title{Effect of ice charging on the astrochemistry of interstellar sulfur-bearing species on amorphous solid water}
   
   \author{T. Vorsselmans\inst{1}\corrauth{tobe.vorsselmans@uantwerpen.be}
        \and I. Grubova\inst{1}\email{Irina.Grubova@uantwerpen.be}
        \and K. Verhagen\inst{1}\email{Kaat.Verhagen@student.uantwerpen.be}
        \and T. Guldentops\inst{1}\email{Tibe.Guldentops@student.uantwerpen.be}
        \and R. Buimer\inst{1}\email{Rianne.Buimer@student.uantwerpen.be}
        \and C. King\inst{1,2}\email{christopher1.king@umconnect.umt.edu}
        \and E. C. Neyts\inst{1}\email{erik.neyts@uantwerpen.be}
        }

   \institute{University of Antwerp, Dept. Chemistry, Research group MOSAIC, Groenenborgerlaan 171, 2020, Antwerp, Belgium
       \and University of Montana, Department of Chemistry, Chemistry Building 115, 32 Campus Drive, Missoula, MT 59812, USA
    }

   \date{Received September 30, 20XX}
 
  \abstract
   {Understanding where up to $99\%$ of the expected sulfur is hidden in dense molecular clouds remains one of the long-standing unresolved problems in astrochemistry. Binding energies (BEs) control desorption, diffusion, and residence times of molecules on amorphous-solid-water (ASW) mantles and hence also determine these input parameters in gas–grain models. The effect of excess negative charge in ASW on sulfur adsorption remains entirely unexplored.}
   {We aim to derive statistically robust and physically interpretable BE distributions for atomic S and the sulfur-bearing molecules H$_2$S, SO$_2$, and OCS on neutral and negatively charged ASW and assess how excess negative charge alters their retention and potential role in the sulfur reservoir of cold, dense molecular clouds.}
   {The basis set superposition error (BSSE) and zero-point energy (ZPE) corrected BEs of S, SO$_2$, OCS, and H$_2$S were calculated via density functional theory (DFT) using the ORCA software. Molecule-specific DFT levels of theory were first selected from benchmark calculations on neutral and charged small water complexes against coupled-cluster reference energies. The selected protocols were then applied to study adsorption on neutral and charged ASW clusters. A range of adsorption sites were sampled on five independent amorphous ice clusters, yielding BE distributions that account for the site heterogeneity of ASW.}
   {Neutral ASW yields broad, site-dependent BE distributions consistent with previous water-ice estimates. On charged ASW, three general cases are identified: the BE always increases due to electron transfer (S-atom, SO$_2$), increases slightly without any electron transfer (H$_2$S), or remains the same unless an electron transfer occurs under specific conditions (OCS). These findings are inherently linked to the molecular properties of these molecules.}
   {Excess negative charge does not uniformly increase sulfur binding on ASW. Instead, it produces molecule-specific adsorption regimes, showing that charged ASW must be treated explicitly in gas–grain models of sulfur chemistry.}


   \titlerunning{Effect of Ice Charging on S-bearing Compounds}
   \maketitle
\nolinenumbers

\section{Introduction}

Interest in understanding the chemical evolution of the interstellar medium (ISM) dates back to the late 1930s and early 1940s, when the unexpected optical detection of CH, CN, and CH$^+$ provided the first unambiguous evidence that space, despite its extremely low densities and temperatures, harbours a chemically active environment \citep{Dunham1937,Swings1937,McKellar1940,Douglas1941}.
However, the field changed decisively only three decades later, when breakthrough advances in observational and spectroscopic capabilities enabled the detection of polyatomic molecules (NH$_3$ and H$_2$O) and the organic molecule formaldehyde (H$_2$CO) in the ISM \citep{Cheung1968,Cheung1969}, overturning the widespread belief that interstellar chemistry would remain limited to diatomics \citep{Snyder2006}.

To date, more than 340 molecular species have been detected in the ISM and circumstellar medium (CSM; \citealt{Mondal2026,McGuire2022}). This inventory is dominated by acyclic organic molecules or radicals with carbon-chain backbones, whereas cyclic species remain comparatively rare. The wide variety of organic molecules detected so far includes aldehydes, alcohols, acids, amines, and carboxamides, that is, several of the principal functional groups from which prebiotic chemistry can emerge. The presence of such chemical diversity, together with the detection of increasingly complex species, indeed leads to a clear conclusion: gas-phase chemistry alone cannot account for the observed molecular richness of the ISM \citep{Herbst2009,Ceccarelli2017}.

A significant step forward in our understanding of chemistry in the ISM came with the realisation that, in cold and dense interstellar clouds, the observed molecular inventory is inseparable from the physics and chemistry of dust grains. It is well known that molecular clouds are the coldest ($\sim 10 \text{K}$) and densest ($10^4 - 10^6 \text{cm}^{-3}$) regions of the ISM and are of particular importance because they are the sites where stars and planets form. In these environments, dust grains with carbonaceous or silicate cores become coated by thick, water-rich icy mantles, generally described as amorphous solid water (ASW; \citealt{Boogert2015,Hama2013}). These icy mantles are not passive condensates, but chemically active interfaces that can concentrate reactants, catalyse reactions by lowering activation barriers, and dissipate excess energy released in exothermic processes \citep{Cuppen2017,Minissale2022,Ceccarelli2023}. In this context, the binding energy (BE) of a species to an ASW surface is not a secondary surface parameter, but one of the key quantities that determines whether a molecule resides on the ice surface, undergoes surface diffusion, participates in surface reactions, or desorbs into the gas phase. However, recent work has also demonstrated that adsorption on ASW cannot generally be represented by a single characteristic such as the BE. Because ASW is structurally heterogeneous, it presents a manifold of local binding environments, so the adsorption energy depends on the site, the molecular orientation, and the surrounding H-bond pattern \citep{Ceccarelli2023,Ferrero2020,Perrero2022a,Tinacci2022,Tinacci2023}. For astrochemical modelling, this is a fundamental consideration; adsorption, desorption, and diffusion rates depend exponentially on the BE, which means that even moderate errors or oversimplified single-value representations can propagate into large errors in predicted abundances, gas--grain partitioning, and snow-line locations \citep{Cuppen2017,Minissale2022,Ceccarelli2023,Ferrero2020,Perrero2022a,Tinacci2022,Tinacci2023,Bariosco2024}.

The presence of sulfur, the tenth most abundant element in the Universe and one of the six CHNOPS elements essential to terrestrial biochemistry, is to date a puzzling issue in astrochemistry. In diffuse clouds, sulfur is only mildly depleted and remains close to its cosmic abundance, whereas in translucent and dense clouds its observable gas-phase abundance drops by one to several orders of magnitude \citep{Rodriguez-Baras2021,Vidal2017,Laas2019}. The problem does not disappear in warmer sources. Even in hot cores and hot corinos, where water-rich icy mantles have sublimated, the sum of detected gaseous sulfur-bearing species still falls far short of the elemental sulfur abundance \citep{Blake1994,Wakelam2004}. This long-standing mismatch is well known as the sulfur-depletion problem; sulfur is not absent from the ISM, but a substantial fraction of it must be hidden in observationally elusive, volatile, semi-refractory, or refractory reservoirs \citep{Rodriguez-Baras2021,Vidal2017,Laas2019,Blake1994,Wakelam2004}.

The most straightforward grain-surface expectation is that atomic sulfur, once accreted onto cold grains, should hydrogenate efficiently to H$_2$S, in close analogy with the formation of water from atomic oxygen \citep{Caselli1994}. However, this simple picture fails observationally. To date, OCS is the only sulfur-bearing molecule securely detected in interstellar ices, while SO$_2$ remains at best tentative; both are far too scarce to close the sulfur budget \citep{Palumbo1995,McClure2023,Palumbo1997,Sturm2023,Boogert2022}. H$_2$S, despite being the obvious hydrogenation product and a widely discussed candidate for the major sulfur reservoir, has still not been securely identified in interstellar ices, even in recent JWST-era studies \citep{McClure2023,Boogert2022}. The paradox becomes even more apparent when compared with Solar System relics: in comet 67P/Churyumov–Gerasimenko, the summed sulfur inventory approaches the solar sulfur abundance, H$_2$S is one of the dominant sulfur carriers, and ammonium hydrosulfide salts have also been identified \citep{Calmonte2016,Biver2021,Altwegg2022}. Sulfur chemistry on grains is therefore clearly efficient, but the retention, transformation, and release of sulfur-bearing species in interstellar ices are more intricate than a simple freeze-out scenario would suggest.

In addition, laboratory studies have further revealed the complexity of this chemistry. Irradiation and photoprocessing of H$_2$S-containing ices repeatedly yield H$_2$S$_2$, polysulfanes, sulfur allotropes, OCS, CS$_2$, and refractory sulfur-rich residues \citep{Jimenez-Escobar2011,Jimenez-Escobar2014,Cazaux2022,Herath2025}. However, the same experiments also demonstrate the limits of the currently available diagnostics. FTIR spectra suffer from overlapping bands, quadrupole mass spectrometry can fragment products in the ion source, and a non-negligible fraction of the sulfur chemistry remains unidentified \citep{Herath2025}. Moreover, temperature-programmed desorption experiments provide essential constraints on BEs, but the extracted values depend on substrate morphology, surface coverage, pre-desorption diffusion, and the assumptions underlying the conversion of desorption traces into BE values \citep{Brown2007,Chaabouni2018}. Although experiments are essential, they do not provide a complete microscopic picture of sulfur adsorption on realistic interstellar ices. Hence, atomistic modelling has become a widely used complement to experiments. Over the past decade, theoretical studies have progressed from single-water proxies and small clusters to periodic models, large amorphous clusters, ONIOM approaches, and hybrid methods combining quantum mechanics with molecular mechanics (QM/MM) \citep{Ferrero2020,Perrero2022a,Tinacci2022,Tinacci2023,Bariosco2024,Duflot2021,Bovolenta2022,Sameera2017,Perrero2022b}. These studies consistently point to three conclusions relevant to the present work. First, the realism of the ice model has a strong effect on the computed BEs. Second, amorphous substrates produce a distribution of BE values rather than a single characteristic value. Third, sulfur-bearing species require careful treatment, as dispersion interactions, molecular polarisability, and local-site geometry all contribute to the adsorption energetics \citep{Perrero2022a,Bariosco2024}.
However, nearly all computational studies have assumed an electronically neutral ice surface, a convenient approximation that may not be physically realistic. Interstellar grains are continuously exposed to cosmic rays, energetic particles, and low-energy secondary electrons generated during the irradiation of condensed matter. As a result, grain charging is regarded as an intrinsic aspect of interstellar grain microphysics rather than an exceptional case. In the context of sulfur chemistry, the inclusion of grain charge in astrochemical models was first implemented to account for the enhanced accretion of gas-phase S$^+$ ions onto negatively charged dust grains via Coulomb attraction \citep{Ruffle1999,Fuente2019,Cazaux2022,Fuente2023}. While these model-based approaches demonstrate the importance of grain charging for sulfur depletion, they did not consider how excess charge modifies the BE of adsorbed species. More recently, \citet{Perrero2024a} investigated the adsorption of neutral and charged species on bare silicate nanoclusters, demonstrating that substrate charge significantly changes adsorption energetics. Nevertheless, the influence of excess negative charge specifically on ASW ice mantles, where most sulfur species are expected to reside in dense clouds, remains unexplored.

In our previous study of small interstellar molecules adsorbed on neutral and negatively charged ASW, we showed that the effect of surface charging is strongly species-dependent; it can enhance adsorption for some molecules, leave it largely unchanged for others, and weaken it for yet others \citep{Vorsselmans2025}. This raises the question of whether sulfur-bearing adsorbates, which span a wide range of electronic properties, respond to charged ASW in a similarly non-uniform manner.

Here, we present a computational study of the adsorption of S, H$_2$S, SO$_2$, and OCS on neutral and singly negatively charged ASW using density functional theory (DFT). We discuss the differences in BEs for the studied molecules with statistical significance in detail and compare calculated BE values with experimental and/or theoretical data reported in the literature. These results have direct implications for astrochemical modelling and could shed new light on sulfur depletion.

\section{Method}

Five ASW clusters were created for the BE calculations, ensuring a large range of unique binding sites. Packmol v21.0.2 \citep{Martinez2009} was used to create these surfaces by placing 30 water molecules in a sphere with a $6.0 {\AA}$ radius. ORCA 6.0 \citep{Neese2022} was used for all subsequent calculations. All five ice clusters were pre-optimised before adding any molecules using B3PW91/D3BJ/def2-TZVPPD at the TightSCF OPT level \citep{becke1988,becke1993,perdew1992,grimme2010,grimme2011,weigend2005,rappoport2010}. A frequency calculation was performed on the optimised cluster to ensure that, after optimisation, the randomly generated cluster is stable, i.e. there are no imaginary frequencies found. If necessary, the optimisation criteria were tightened to TightOpt and the numerical integration grid was refined to DefGrid3 to resolve any imaginary frequencies. Once all clusters were optimised in the neutral state, the charge and multiplicities were changed to $-1$ and $2$, respectively, to create a charged ASW cluster. Again, the charged surface was optimised, and a frequency calculation was performed analogously to the neutral cluster until geometric convergence without any imaginary frequencies.

Before any calculations were made, a benchmark study was performed to identify the best functional, basis set, and dispersion correction combination for each molecule. The benchmark was performed by calculating the BE of a molecule on a water dimer for a range of functionals, basis sets, and dispersion corrections and was tested against the CCSD(T)/aug-cc-pVTZ level of theory \citep{purvis1982,raghavachari1989,dunning1989,kendall1992}. The full benchmark can be found in Tables \ref{table:benchmark_H2S}--\ref{table:benchmark_SO2}. Not only are the BEs on both the neutral and charged ice clusters important, the difference between these two values is especially noteworthy; thus, all three values were taken into consideration when choosing the best functional, basis set, and dispersion correction combination. Table \ref{table:method} shows the conclusion from the benchmark study. Thus, we employed a distinct functional, basis set, and dispersion correction combination for every specific molecule to ensure optimal accuracy per molecule. This implies that while we cannot directly compare BEs between molecules, we obtained reliable BE differences per molecule on a charged ice cluster versus on a neutral ice cluster.

\begin{table}
\caption{Selected functional, basis set, and dispersion correction combination for every molecule.}
\label{table:method}
\centering
\begin{tabular}{c c c c}
\hline\hline
Species & Functional & Basis set & Dispersion \\
\hline
    S & RevPBE38 & Def2-TZVPPD & D3(0) \\
    H$_2$S & B3PW91 & Def2-TZVPPD & D3(BJ) \\
    SO$_2$ & RevPBE38 & Def2-TZVPPD & D3(0) \\
    OCS & $\omega\text{B97M}$ & Def2-TZVPPD & D4Rev \\
\hline
\end{tabular}
\tablefoot{Combination selected according to the benchmark against CCSD(T)/aug-cc-pVTZ.}
\end{table}

For every ASW cluster, six binding sites were created by simply taking the positive and negative sides of the x-, y-, and z-axes in a Cartesian coordinate system. The starting configurations were thus systematically created without any consideration for the exact starting binding site to provide a random coverage of the whole cluster. Using this strategy rather than manually selecting binding sites, statistical insight into the BEs was obtained without premeditated bias in the setup of our calculations. This approach corresponds to a stochastic hit-and-stick process of molecules landing on a surface. Consistently with this approach, BE  averages were calculated as equal-weight averages (thus assuming no surface diffusion) rather than Boltzmann-weighted averages (which would assume diffusion and thermal redistribution of molecules over the available binding sites).

The molecules to be placed on the ASW cluster are H$_2$S, SO$_2$, S, and OCS. These were chosen to have a diverse group of molecular properties, where H$_2$S can act as both an H-bond donor and acceptor, and SO$_2$, S, and OCS act only as H-bond acceptors. The multiplicities of all molecules are singlet, but the S atom in its ground state contains two unpaired electrons, changing the multiplicity to a triplet state. For the neutral complexes (ASW cluster with a molecule), a zero and one charge and multiplicity were employed for H$_2$S, SO$_2$, and OCS. For S, zero and three charges and multiplicities were necessary. In the charged state, all complexes were assigned a charge of $-1$ and a multiplicity of 2.

The BE was defined as the negative of the interaction energy, such that positive values correspond to stable adsorption and larger BEs indicate stronger molecule–surface interactions:
\begin{equation} \label{eq:BE}
    BE = [(E_{molecule} + E_{surface}) - E_{complex}] + \Delta E_{BSSE} + \Delta E_{ZPE}
.\end{equation}
Here, $E_{molecule}$, $E_{surface}$, and $E_{complex}$ denote the total energies of the isolated adsorbate, the bare ASW cluster, and the optimised adsorption complex, respectively. The basis set superposition error (BSSE) was corrected using the counterpoise scheme of \citet{boys1970}. Zero-point energy (ZPE) corrections were obtained from harmonic vibrational frequencies and added to the electronic binding energies. The reported BEs therefore include both BSSE and ZPE corrections.

\section{Results}
\subsection{Benchmark and ZPE correction}

The results from our benchmark are presented in Tables \ref{table:benchmark_H2S}--\ref{table:benchmark_SO2}. For each molecule, we calculated the BE of each of the four molecules on a water dimer, testing a range of functional, basis set, and dispersion correction combinations by comparison to CCSD(T)/aug-cc-pVTZ. For each molecule, we took into consideration the error of the neutral BE ($\text{Error}^0$), the error of the charged BE ($\text{Error}^{-1}$), and the error of the change in BE ($\text{Error}^\Delta$). For H$_2$S, we found that B3PW91/def2-TZVPPD/D3(BJ) gave the best correspondence to CCSD(T) with $-1.15\%$ $\text{Error}^0$, $-3.46\%$ $\text{Error}^{-1}$, and $-5.07\%$ $\text{Error}^\Delta$. The B3PW91 functional slightly underestimates the BE for both neutral and charged complexes. The error of the BE values is $-14$ K and $-110$ K for the neutral and charged complexes, respectively, which is well within chemical accuracy ($< 500 \text{K}$). Next, the OCS benchmark gave the $\omega\text{B97M}$-D4REV/def2-TZVPPD as the best option to calculate the BE \citep{mardirossian2016,caldeweyher2019}. Especially for the neutral complex, it showed an $\text{Error}^0$ of only $-0.14\%$, corresponding to an underbinding of $3 \text{K}$. The charged BE was also good, with a slight overbinding of $57 \text{K}$, corresponding to an $\text{Error}^{-1}$ of $3.79\%$. The change in BE results in an $\text{Error}^\Delta$ of $-22.57\%$, which appears large but, in absolute terms, corresponds to a difference of only $-59 \text{K}$ compared to the BE change obtained at the CCSD(T) level. Lastly, both S and SO$_2$ have RevPBE38/def2-TZVPPD/D3(0) as the best performing combination \citep{zhang1998,santra2012}. For S, this combination showed the smallest error for both the neutral and charged BE, with overbindings of $8.57\%$ $\text{Error}^0$ ($149 \text{K}$) and $3.50\%$ $\text{Error}^{-1}$ ($1330 \text{K}$). The change in BE showed the second smallest error of $3.25\%$ $\text{Error}^\Delta$ ($1181 \text{K}$). Finally, for SO$_2$, all functionals overestimated the BE on the neutral dimer. Here, RevPBE38 gave the smallest error of $17.24\%$ $\text{Error}^0$ ($627 \text{K}$). This is slightly above what would be considered chemical accuracy ($> 500 \text{K}$), but it is the best-performing functional we found, and both the charged BE and the change in BE were calculated accurately with $0.23\%$ $\text{Error}^{-1}$ ($69 \text{K}$) and $-2.16\%$ $\text{Error}^\Delta$ ($-558 \text{K}$).

Once the best method for each molecule was selected based on our benchmark, the method was tested against the literature. The comparison was done with the BEs of the molecules on a tetramer water cluster, as done by \citet{Das2018}. To accomplish this, the four molecules were placed on a water tetramer and optimised using the best method according to Table \ref{table:method}. Then the BEs were calculated both at the DFT level and at the higher DLPNO-CCSD(T1)/aug-cc-pVTZ level \citep{riplinger2013,liakos2015}. Table \ref{table:Tetramer} shows the results comparing our DFT method with both the literature and DLPNO-CCSD(T1). It can be seen that our DFT for each method shows good correspondence with both the DLPNO-CCSD(T1) level and the literature value. It is interesting to note here that the BE of OCS on a charged tetramer shows a different behaviour than on the water dimer. On the dimer, the BE was not affected by the charging of the water cluster, while here on the tetramer there is a significant increase of an order of magnitude when the tetramer is charged. We found that the origin of this discrepancy lies in the electron-accepting behaviour of OCS. On the dimer, the OCS molecule refused to accept the extra electron since it requires a significant structural reorganisation to form the OCS$^-$ anion. It seems that on a tetramer it is beneficial for the complex to overcome this reorganisation and form the OCS$^-$ anion, leading to a significant increase in BE. This interpretation is consistent with previous gas-phase cluster studies showing that isolated OCS$^-$ is metastable or only weakly bound, whereas hydration can stabilise OCS$^-$ H$_2$O and larger OCS$^-$ (H$_2$O)$_k$ cluster anions \citep{Surber2002}.

\begin{table}
\caption{Comparison of the selected DFT protocols with DLPNO-CCSD(T1)/aug-cc-pVTZ and literature for a molecule on an H$_2$O tetramer.}
\label{table:Tetramer}
\centering
\begin{tabular}{C{3.9em} C{3.9em} C{3.9em} C{3.9em} C{3.9em}}
\hline\hline
Species & Charge & DFT & DLPNO-CCSD(T1) & Das et al. (2018)$^{(1)}$ \\
\hline
    \multirow{2}{3.9em}{\centering S} & $0$ & $1163$ & $1162$ & $1428$ \\
      & $-1$ & $43031$ & $40116$ & - \\
    \hline
    \multirow{2}{3.9em}{\centering H$_2$S} & $0$ & $2178$ & $2149$ & $2556$ \\
           & $-1$ & $1803$ & $1864$ & - \\
    \hline
    \multirow{2}{3.9em}{\centering SO$_2$} & $0$ & $3480$ & $3465$ & $3745$ \\
           & $-1$ & $35552$ & $32532$ & - \\
    \hline
    \multirow{2}{3.9em}{\centering OCS} & $0$ & $1172$ & $1480$ & $1571$ \\
        & $-1$ & $20209$ & $16194$ & - \\
\hline
\end{tabular}
\tablefoot{All BEs are given in Kelvin.}
\tablebib{$^{(1)}$\citet{Das2018}.}
\end{table}

We stress that a benchmark as performed here is always somewhat arbitrary. First, it is not feasible to test every possible functional, basis set, and dispersion correction combination. Second, the choice of the reference system---here the dimer with an adsorbed molecule---is of importance. On larger ice clusters, the balance of the various types of interactions between the molecule and the cluster change, which is already the case for OCS on a charged tetramer versus a charged dimer. Clusters larger than a few molecules are impractical at the high CCSD(T) level of theory. Nevertheless, we show here that the neutral BEs are accurate when compared to both CCSD(T) and literature values for both a dimer and tetramer water cluster. Since the change in BE is also consistent with CCSD(T) on a water dimer, we are confident that the presented method is robust enough to be used as a proof of concept to investigate the potential change in BE of sulfur-containing compounds on charged ice surfaces.

The ZPE correction was accounted for by multiplying the non-corrected BE with a coefficient of one. This coefficient was determined by a linear regression between the non-ZPE-corrected and ZPE-corrected BEs for a test sample of species. From this regression, it was found that the ZPE correction was negligible at high BE regimes, while at lower BE regimes the ZPE correction was significant. Hence, only the low BEs were ZPE-corrected differently based on two categories: neutral BEs (ratio = $0.762$) and low charged BEs (ratio = $0.748$).

 \begin{figure*}
\centering
   \includegraphics[width=17cm]{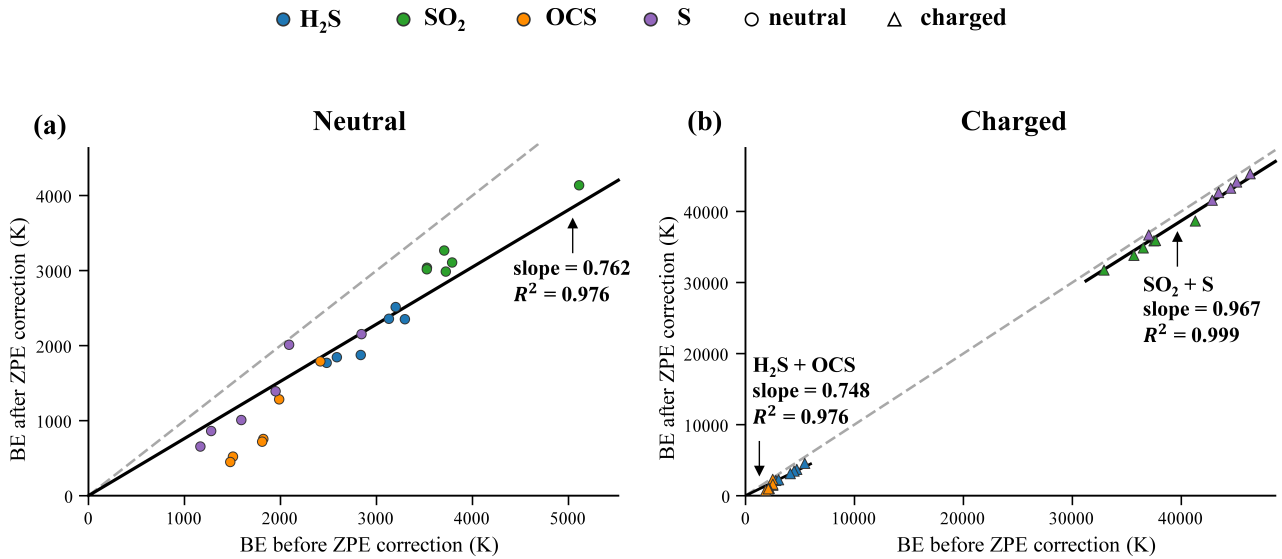}
     \caption{Linear fits between the BE values before and after ZPE correction for H$_2$S, SO$_2$, OCS, and S on ASW. (A) Neutral ASW cluster. (B) Charged ASW cluster. The dashed grey line indicates y=x, whereas the black lines show linear fits through the origin. A single fit was obtained for the neutral systems, while two separate fits were used for the charged low-BE (H$_2$S + OCS) and high-BE (SO$_2$ + S) regimes. In these fits, the BE is BSSE-free.}
     \label{fig:ZPE}
\end{figure*}

\begin{figure*}
    \centering
    \includegraphics[width=\textwidth]{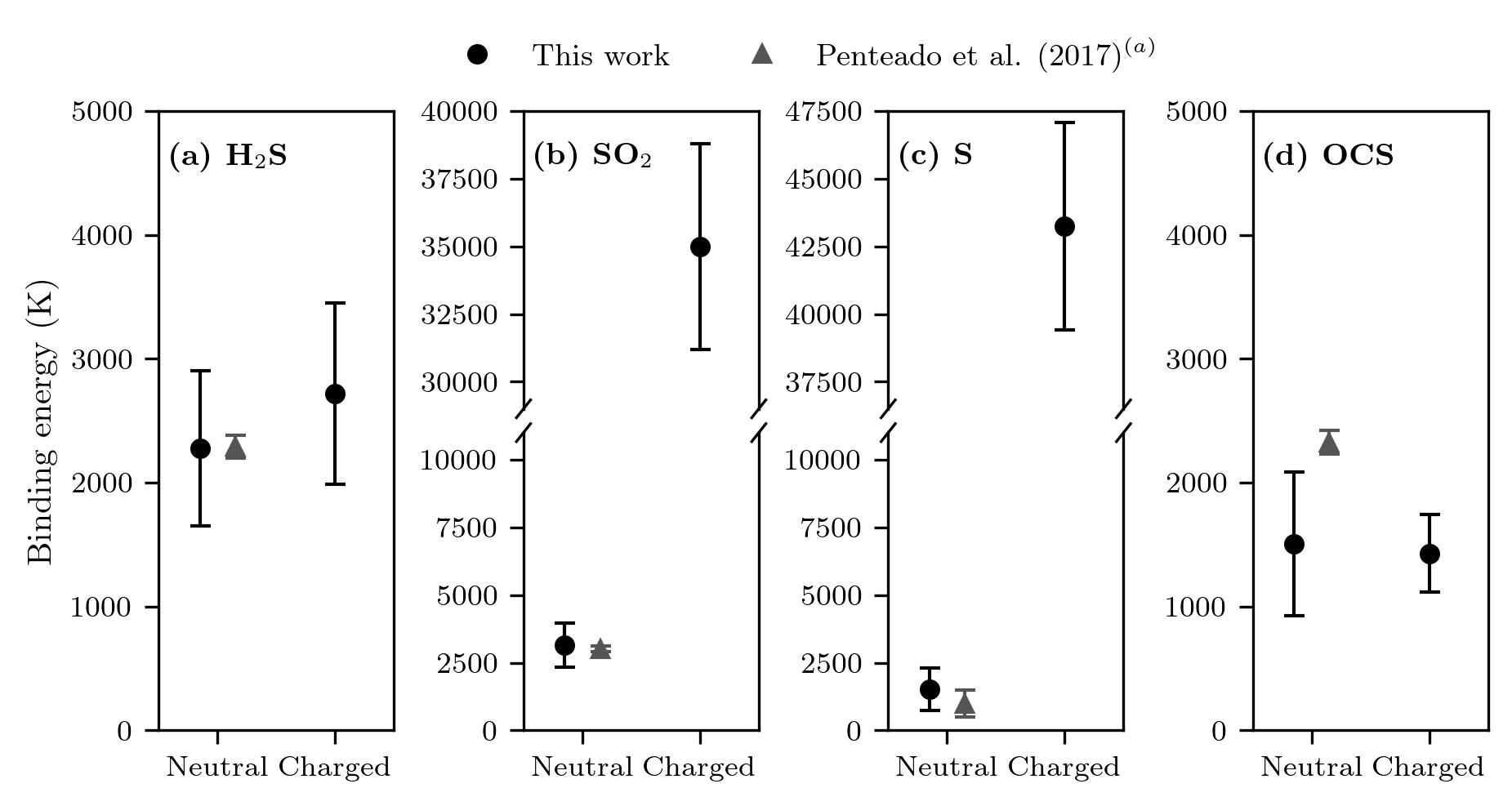}
    \caption{Binding-energy distributions for (a) H$_2$S, (b) SO$_2$, (c) S-atom, and (d) OCS on neutral and charged ASW. The mean value is shown as a marker, while the error bars show the standard deviation for our samples. The literature values from $^{(a)}$\citet{Penteado2017} are recommended values, with the uncertainties as mentioned in that paper.}
    \label{fig:results}
\end{figure*}

The resulting neutral and charged BE distributions are summarised in Fig. \ref{fig:results} and Table \ref{table:results}. We next discuss the four adsorbates separately to relate their charge response to the underlying binding motifs and electron-transfer behaviour.

\begin{table*}
\caption{DFT calculations for all studied species on the ASW surface.}
\label{table:results}
\centering
\begin{tabular}{c c c c c c c c c}
\hline\hline
Species & Charge & Min BE (K) & Max BE (K) & $\mu_{BE}$ (K) & $\sigma_{BE}$ & Disp & N & p-value ($\mu_{BE}$ equal?) \\
\hline
    \multirow{2}{4em}{\centering H$_2$S} & $0$ & $937$ & $3500$ & $2276$ & $625$ & $57\%$ & $30$ & \multirow{2}{8em}{\centering $0.015$ (No)} \\
      & $-1$ & $1399$ & $4553$ & $2717$ & $732$ & $53\%$ & $30$ & \\
    \hline
    \multirow{2}{4em}{\centering SO$_2$} & $0$ & $1762$ & $5139$ & $3145$ & $819$ & $61\%$ & $30$ & \multirow{2}{8em}{\centering $1.03\text{x}10^{-46}$ (No)} \\
           & $-1$ & $25527$ & $42113$ & $34992$ & $3803$ & $8\%$ & $30$ & \\
    \hline
    \multirow{2}{4em}{\centering S} & $0$ & $552$ & $3254$ & $1571$ & $728$ & $69\%$ & $30$ & \multirow{2}{8em}{\centering $2.93\text{x}10^{-53}$ (No)} \\
           & $-1$ & $34650$ & $49895$ & $43250$ & $3836$ & $4\%$ & $30$ & \\
    \hline
    \multirow{2}{4em}{\centering OCS} & $0$ & $889$ & $3621$ & $1504$ & $580$ & $82\%$ & $22$ & \multirow{2}{8em}{\centering $0.577$ (Yes)}\\
        & $-1$ & $882$ & $1956$ & $1429$ & $315$ & $81\%$ & $26$ & \\
\hline
\end{tabular}
\tablefoot{P-value indicates whether the mean values of the neutral and charged cases are statistically equal. Min BE: minimum binding energy; Max BE: maximum binding energy; $\mu_{\text{BE}}$: mean binding energy; $\sigma_{\text{BE}}$:  standard deviation; Disp: dispersion correction averaged over all samples; N: sample size.}
\end{table*}

\subsection{H$_2$S}

The BE of H$_2$S on the charged surface ($2717 \pm 732 \text{K}$) is slightly higher than on the neutral surface ($2276 \pm 625 \text{K}$), corresponding to an increase of $441 \text{K}$ ($+19.4\%$) (Fig. \ref{fig:results}a and Table \ref{table:results}). A Welch t test yielded p = $0.015$ ($< 0.05$), indicating a statistically significant difference between the mean values at this level of significance. Nevertheless, the small difference in mean values and the fact that the standard deviations of both groups are larger than this difference show that many data points in both groups overlap.

\begin{figure*}
    \centering
    \begin{subfigure}[b]{0.48\textwidth}
        \centering
        \includegraphics[width=0.5\textwidth]{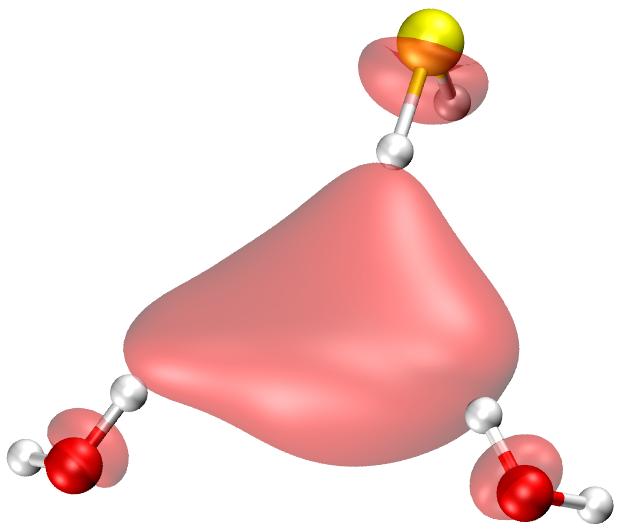}
        \caption{}
        \label{fig:H2S_a}
    \end{subfigure}
    \hfill
    \begin{subfigure}[b]{0.48\textwidth}
        \centering
        \includegraphics[width=0.7\textwidth]{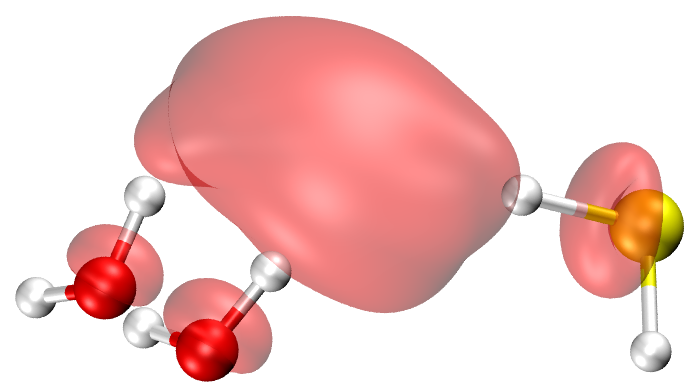}
        \caption{}
        \label{fig:H2S_b}
    \end{subfigure}
    \caption{Spin-density plot of H$_2$S on a charged surface. The red lobe demonstrates where the extra electron resides within the complex. In both (a) and (b), the extra electron is located in a pocket created by the coordination of nearby H atoms from both the ASW surface and the H$_2$S molecule.}
    \label{fig:H2S}
\end{figure*}

For the neutral BEs, we observe that the lowest BEs are found when H$_2$S interacts with regions of the surface that lack dangling H- or O-atoms, leading to a low BE ($937 \text{K}$). When a H atom dangling from the surface (denoted as dH) interacts with the S atom (i.e. when a dH--S interaction is established), a significantly higher BE is found ($1849 \text{K}$), i.e. a difference of $912 \text{K}$ compared to the case where no interactions with dH or O are present. Further increases in BEs are due to S-H--O interactions. This demonstrates that the positioning and orientation of the water molecules around the H$_2$S molecule is of great importance. When the S-H--O interactions can be formed without hindrance from other H atoms from the ASW, the BE increases. The repulsion between the partially positive H atoms on the H$_2$S molecule and the ASW is thus very important in these higher BE configurations.
For the charged BEs, we observe the same trend as for the neutral BEs. Once a dH--S interaction is established, the BE increases. However, there are some configurations that interact with the extra electron alongside water, as shown in Fig. \ref{fig:H2S}. The lowest BE found ($1399 \text{K}$) is a direct interaction between an H atom from H$_2$S and the extra electron on the surface. This configuration is shown in Fig. \ref{fig:H2S_a} and yields a low BE since there is only one interaction between the H$_2$S molecule and the surface through this loosely bound electron, which is also interacting with two dH atoms from the ASW surface. Thus, the electron is localised in the pocket formed by the three H atoms and determines the BE of the H$_2$S molecule on the surface. The next interesting interaction we find between H$_2$S and the extra electron, shown in Fig. \ref{fig:H2S_b}, has a BE of $4025 \text{K}$. This value is already significantly higher than that of the average BE on a neutral surface ($2276 \text{K}$). However, the maximum value on a charged surface of $4553 \text{K}$ actually shows no interaction with the extra electron on the surface.

Although the neutral-to-charged BE shift for H$_2$S is statistically significant within the present dataset, its magnitude is smaller than the spread among literature estimates for neutral water ice (see Table \ref{table:literature}). This indicates that the H$_2$S charge effect should be interpreted cautiously; grain charging induces only a moderate stabilisation within the same ASW model and sampling protocol rather than driving a distinct charge-dominated adsorption regime.

\subsection{SO$_2$}

The BE for SO$_2$ on the charged surface ($34992 \pm 3803 \text{K}$) is an order of magnitude higher than on the neutral surface ($3145 \pm 819 \text{K}$), corresponding to an increase by a factor of more than ten (Fig. \ref{fig:results}b, Table \ref{table:results}). Welch’s t test yields p = $1.03\text{x}10^{-46}$ ($< 0.05$), indicating a statistically significant difference between the mean values.

\begin{figure*}
    \centering
    \begin{subfigure}[b]{0.48\textwidth}
        \centering
        \includegraphics[width=0.4\textwidth]{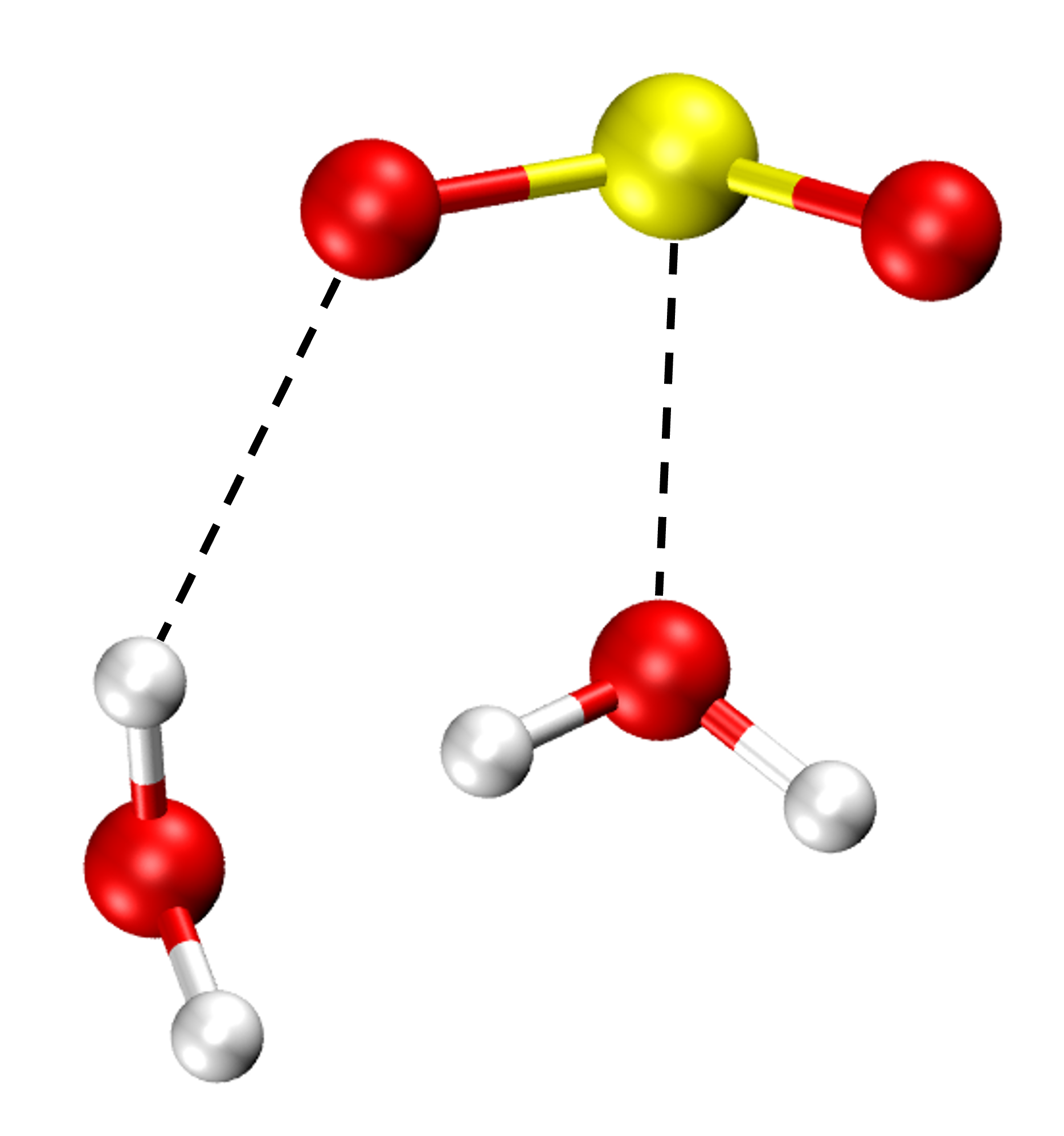}
        \caption{}
        \label{fig:SO2_a}
    \end{subfigure}
    \hfill
    \begin{subfigure}[b]{0.48\textwidth}
        \centering
        \includegraphics[width=0.5\textwidth]{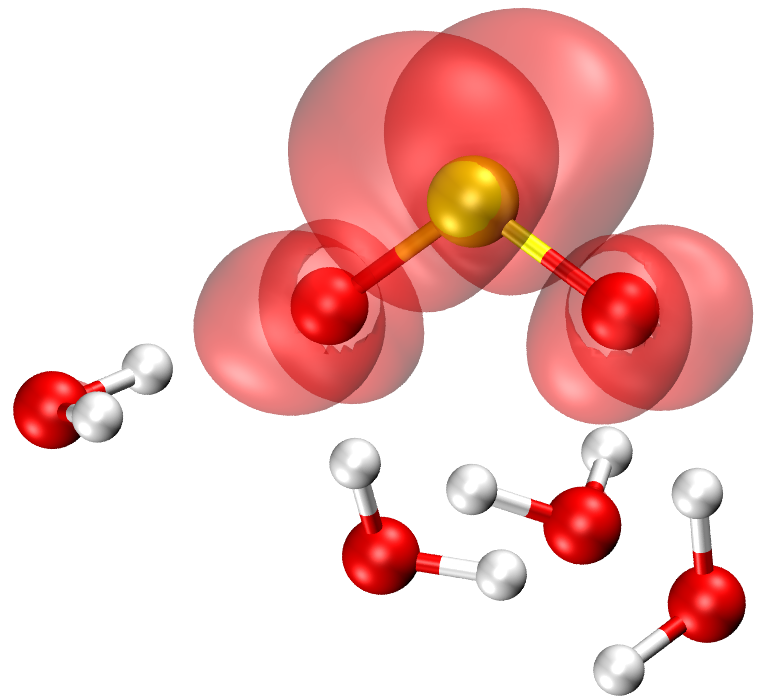}
        \caption{}
        \label{fig:SO2_b}
    \end{subfigure}
    \caption{SO$_2$ on a (a) neutral surface and a (b) charged surface. Panel (a) shows both the S--O interaction and the dH--O interaction often found in these neutral ASW-SO$_2$ complexes. The red lobe in panel (b) shows the spin density of the charged ASW-SO$_2$ complex. The extra electron clearly resides in the LUMO of SO$_2$ and interacts with the surface through multiple dH--O interactions.}
    \label{fig:SO2}
\end{figure*}

In the SO$_2$ molecule, the S atom is partially positively charged. Because of this, we observe that the SO$_2$ forms S--O interactions with the ASW. Additionally, both dH--O and H--O interactions are established between dangling H atoms (i.e. H atoms that are part of the H-bonded network) and an O atom of SO$_2$. In Fig. \ref{fig:SO2_a}, the S--O interaction and dH--O interaction are shown to illustrate these interactions between SO$_2$ and the neutral ASW cluster. As is the case for the H$_2$S molecule, the configuration of the water around the binding site is very important for the strength of the interaction. The highest reported BE values are for configurations where SO$_2$ can form a dH--O interaction with one of its O atoms and a H--O interaction with its other O atom, apart from the S--O interaction. This configuration thus involves three interactions between the SO$_2$ and the ASW, resulting in a high BE.

When a charge is introduced into the system, we see that the extra electron always resides on SO$_2$, creating the SO$_2^-$ anion, as shown in Fig. \ref{fig:SO2_b}. The S--O interaction observed in the neutral case does not form with the anion since the $\pi^*$ orbital is mostly located on the S atom and now carries a negative charge. The water molecules from the ASW also preferably interact with the SO$_2^-$ anion instead of creating an H-bond network. This can be seen from the number of dH--O interactions formed, which ranges from only one at lower BEs up to four at higher BEs. The BEs reported here for the charged system (average of $34992 \text{K}$) are an order of magnitude higher than the energies for the neutral system (averaging at $3145 \text{K}$), showing a clear change in binding mode when the surface is charged.

\subsection{S}

The BE for an S atom on the charged surface ($43250 \pm 3836 \text{K}$) is significantly higher than on the neutral surface ($1571 \pm 728 \text{K}$), corresponding to an increase by a factor of almost 30 (Fig. \ref{fig:results}c and Table \ref{table:results}). Welch’s t test yields p = $3.72\text{x}10^{-52}$ ($< 0.05$), indicating a statistically significant difference between the mean values.

The most important interactions between the S atom and the ASW cluster are electronic interactions and dispersion forces. The dispersion interactions are important because of the high polarisability of the S atom (Table \ref{table:results}). The average contribution of the Grimme D3 dispersion correction (D3(0)) over all BEs is $69\%$, showing a significant contribution of the dispersion forces. There does seem to be a preference for the S atom to interact with a dH from the surface. From our calculations, placing the S atom at six different places on five different surfaces, the S atom seems to stick wherever we place it near the surface. This is a different behaviour than the other molecules, which show some limited movement over the surface to find an optimal binding site in the vicinity of their original location near the surface. This also explains why there is such a high spread of BEs found for the S atom and, thus, why we have a large standard deviation ($728 \text{K}$) relative to the mean BE ($1571 \text{K}$).

The charged system shows a completely different result. Here, the S atom takes on the extra electron, thereby creating the S$^-$-anion. In the charged case, dispersive forces are nearly negligible, and instead an electrostatic interaction is established between S$^-$ and ASW. Indeed, the Grimme D3 dispersion correction now only accounts for $4\%$ of the total BE on average. The strength of the interaction is purely dependent on the amount of dHs available for interaction with S$^-$. At the lowest value found ($34650 \text{K}$), there are two dH--S$^-$ interactions, while for the highest value ($49895 \text{K}$) there are four dH--S$^-$ interactions observed.

\subsection{OCS}

The BE for OCS on the charged surface ($1429 \pm 315 \text{K}$) is nearly equal to the BE on the neutral surface ($1504 \pm 580 \text{K}$); the difference in mean values is only $75 \text{K}$ ($-5\%$) (Fig. \ref{fig:results}d and Table \ref{table:results}). Welch’s t test yields p = $0.577$ ($> 0.05$), indicating that there is no statistically significant difference between the mean values. Both the neutral and charged complexes are missing some configurations that either did not converge or for which the BE increased to unreasonably high values multiple magnitudes higher than the main results.

Similar to S and SO$_2$, OCS can act as an H-bond acceptor but cannot act as an H-bond donor. However, SO$_2$ has a much larger dipole moment than OCS. The H bonds formed between OCS and the ASW surface are therefore much weaker than those of SO$_2$, as shown in Table \ref{table:results}. In fact, OCS shows the lowest mean BE value ($1504 \text{K}$) of all molecules investigated in this study. The strongest interactions found are indeed dH--O interactions, while the weakest are O--S interactions between the O atoms from the ASW cluster and the S atom from OCS. Indeed, we see that in configurations with low BEs, the OCS-ASW interaction consists mainly of dispersion forces, with a Grimme D4Rev dispersion contribution of $75\%$ to $85\%$ of the total BE. The high dispersion correction is a result of the high polarizability of the S atom in OCS, so it forms a dipole-induced dipole interaction with the O atoms from the cluster. In configurations with higher BEs, the contribution of the dispersion correction drops to $65\%$, signifying a more electrostatic interaction between OCS and the surface through dH--O interactions.

When the system is charged, we do not see any difference in BE with the neutral system, as seen in Table \ref{table:results}. OCS does not readily take on the extra electron as the S-atom or SO$_2$ do, nor does it interact with the loosely bound electron on the surface as H$_2$S does. Since OCS has a lower dipole moment than H$_2$S, it is less likely to interact strongly with the extra electron. The OCS molecule indeed mainly binds to the surface through dispersion interactions with its highly polarisable S atom. However, we did find four cases where an electron transfer to OCS occurs. In these cases, the molecule was initially placed close to the location where the loosely bound electron resides on the charged ASW surface, i.e. within $5 {\AA}$. When an electron transfer occurs, the BE increases significantly---up to $11000 \text{K}$. Since only a few cases were found where the electron transfer occurs, and since it only happens with specific starting conditions before optimisation, we opted to keep these cases out of the general results shown in Table \ref{table:results}. It is, however, important to note that even if it is unlikely that the OCS molecule takes on the extra electron, there are certainly cases where this does happen, resulting in a significant increase in BE.

\section{Discussion}
\subsection{Comparison of neutral BE values with literature data}

Since BEs are a key input for gas–grain astrochemical models, they have been estimated both through experiments and quantum chemical calculations. Here, we put our neutral BE distributions alongside the values reported in the literature (Table \ref{table:literature}). However, the charged ASW results are discussed separately as they probe a charge-driven adsorption regime for which no direct literature values are currently available.

The literature values span a wide range (Table \ref{table:literature}), reflecting differences in ice models, site sampling, and computational or experimental protocols. A direct one-to-one comparison is therefore generally inappropriate. In particular, small clusters reproduce neither the extended H-bond cooperativity nor the site heterogeneity of realistic ASW surfaces \citep{Ferrero2020,Perrero2022a,Bariosco2024}, while desorption-based estimates are sensitive to surface morphology, adsorbate coverage, diffusion, and thermal restructuring of the ice \citep{Minissale2022,Penteado2017}.

For H$_2$S, the neutral BE distribution obtained in this work spans $937-3500$ K, with $\mu = 2276 \text{K}$ and $\sigma = 625 \text{K}$. This range brackets the recommended value of $2290 \pm 90 \text{K}$ \citep{Penteado2017} and overlaps with the single-molecule, cluster, and periodic-ASW values of \citet{Wakelam2017}, \citet{Das2018}, and \citet{Perrero2022a}, respectively. The distribution is, however, shifted to higher values relative to the ACO-FROST large-grain model of \citet{Bariosco2024}, which sampled a much larger population of binding sites and included very weakly bound configurations ($\mu \approx 984 \text{K}$). This comparison suggests that H$_2$S binding is particularly sensitive to the range of adsorption sites sampled. The addition of an excess electron raises the mean BE only moderately, from $2276$ to $2717 \text{K}$, while the neutral and charged distributions remain strongly overlapping. Thus, for H$_2$S, surface charging leads only to a moderate site-dependent stabilisation rather than to a distinct charge-driven adsorption regime.

The neutral SO$_2$ distribution spans $1762-5139 \text{K}$ ($\mu = 3145 \text{K}$, $\sigma = 819 \text{K}$), which is in good agreement with the recommended value of \citet{Penteado2017}, the periodic-ASW distribution of \citet{Perrero2022a}, and the KIDA database entry \citep{Wakelam2015}. Although higher values are reported in UMIST \citep{McElroy2013} and in the H$_2$O-monomer calculations of \citet{Wakelam2017}, our results fall within the main range of previous neutral-ice estimates. The charged-surface case is qualitatively distinct; the excess electron is transferred to SO$_2$, forming SO$_2^-$, and the mean BE rises to approximately $35000 \text{K}$. These values describe the electrostatic stabilisation of an anionic adsorbate within a polar water network and are clearly not comparable with conventional neutral-surface physisorption energies.

Atomic sulfur shows an analogous charge-driven transition. On neutral ASW, the BE values span from $552 \text{K}$ to $3254 \text{K}$, with $\mu = 1571 \text{K}$ and $\sigma = 728 \text{K}$. This range is consistent with the broad uncertainty of the recommended value of \citet{Penteado2017} and overlaps with previous single-water and periodic-ASW estimates \citep{Perrero2022a,Wakelam2017,Perrero2024b}. On neutral ice, the interaction is governed mainly by dispersion and polarisation, since an isolated S atom cannot form conventional H bonds. On negatively charged ASW, the excess electron transfers to the sulfur atom, forming S$^-$, and the mean BE rises to approximately $43000 \text{K}$. This more than an order-of-magnitude increase reflects the transition from weak, dispersion-dominated adsorption to the strong electrostatic stabilisation of the anion.

In contrast to both SO$_2$ and atomic S, OCS behaves differently. The neutral BE distribution, $889-3621 \text{K}$ with $\mu = 1504 \text{K}$ and $\sigma = 580 \text{K}$, overlaps with the lower part of the literature range but lies below most literature central values reported for ASW and database models. This is physically consistent with the molecular character of OCS; it is nearly linear, weakly polar, and can only act as an H-bond acceptor, so its interaction with ASW is largely dominated by dispersion involving the polarisable sulfur atom \citep{Perrero2022a}. In the charged case, the addition of an excess electron does not significantly alter the sampled OCS distribution ($\mu = 1429 \text{K}$; Welch’s t test: p = $0.577$), indicating that OCS does not generally localise the charge under the present conditions. The rare electron-transfer cases should therefore be treated separately from the main neutral/charged comparison.
Thus, the four adsorbates reveal a molecule-dependent response governed by the balance among H-bonding, polarity, dispersion, and the capacity for charge localisation. Overall, the neutral BEs obtained here fall within the broad range of literature values for ASW models, whereas the effect of excess negative charge is highly species-specific. This demonstrates that charge cannot be represented by a uniform scaling of neutral BEs but must be treated explicitly for sulfur-bearing adsorbates.

\subsection{Electron-transfer behaviour}

When a molecule adsorbs on a charged ASW surface, three scenarios are possible: the electron may always transfer to the molecule, the electron may never transfer to the molecule, or the electron transfer may occur only under specific conditions. The molecules discussed in this paper cover all three scenarios. The S atom and SO$_2$  always accept the electron, H$_2$S never accepts the electron, and for OCS the electron transfer is proximity-dependent. The reasons behind the behaviour of the different molecules are threefold. First, there is the thermodynamic driving force, which is defined by the electron affinity of the molecule. Next is the kinetic barrier, which is the reorganisation energy required to accept an electron. Third, when an electron transfer is impossible, the electrostatic properties of the molecule---such as dipole moment and polarisability---dictate the BE. Table \ref{table:transfer} shows these molecular properties, calculated using the method described in the methods section, for the four molecules presented in this paper.

\begin{table}
\caption{Molecular properties associated with the electron-accepting capabilities of the studied molecules.}
\label{table:transfer}
\centering
\begin{tabular}{C{3em} C{3em} C{3em} C{3em} C{3em} m{3em}}
\hline\hline
Species & $EA_{vertical}$ (eV) & $EA_{adiabatic}$ (eV) & $E_{structure}$ (eV) & $\mu$ (D) & e$^-$ transfer? \\
\hline
    H$_2$S & $-1.02$ & $-1.02$ & $0.00$ & $1.00$ & No \\
    OCS & $-0.75$ & $-0.03$ & $0.72$ & $0.70$ & Proximity Dependent \\
    S & $1.99$ & $1.99$ & $0.00$ & $0.00$ & Yes \\
    SO$_2$ & $0.95$ & $1.29$ & $0.34$ & $1.66$ & Yes \\
\hline
\end{tabular}
\tablefoot{$EA_{vertical}$: vertical electron affinity; $EA_{adiabatic}$: adiabatic electron affinity; $E_{structure}$: structural reorganisation energy; $\mu$: dipole moment; e$^-$: electron.}
\end{table}

The S atom and SO$_2$ have a high electron affinity, meaning that electron capture is highly favourable, regardless of where on the ASW cluster they are placed. SO$_2$ shows a small structural reorganisation due to the bond angle narrowing slightly (from $\sim120^\circ$ to $\sim115^\circ$) to accommodate the extra electron. H$_2$S has a highly negative electron affinity; thus, the H$_2$S$^-$ anion is highly unstable and prevents H$_2$S from ever accepting the extra electron from the surface. Since H$_2$S does not accept the electron, it falls back to its electrostatic properties, interacting with the surface with its moderate dipole moment and high polarisability. For OCS, we see a negative vertical electron affinity and an almost zero adiabatic electron affinity. The OCS molecule therefore does not accept the electron in its neutral geometry ($EA_{vertical}$). In its anionic geometry ($EA_{adiabatic}$), however, it can accommodate the extra electron at the cost of a high structural reorganisation ($E_{structure}$). The necessity of OCS to first bend significantly before accepting an electron makes it proximity-dependent. The intense, localised electric field of the loosely bound electron can only trigger the bending of the molecule when the OCS adsorbs close to it. When the OCS adsorbs farther away from the electron, there is no external force to overcome this kinetic barrier.

\subsection{Astrophysical implications}

It has long been recognised that molecules that seem to be missing from, or appear at very low abundances in, the gas phase of cold molecular clouds may be locked up in interstellar ice mantles. The present results support this possibility by virtue of higher-than-expected BEs on charged ice mantles. Our computations show that the BE of H$_2$S slightly increases on a negatively charged surface. This small increase might seem insignificant, but it has a big effect on the desorption rate. At the low temperatures ($\sim10 \text{K}$) of dense molecular clouds, the exponential relation of the desorption rate on the BE dominates ($k_{des} \propto exp(-E_{bind}/T)$) \citep{Minissale2022}. This causes H$_2$S to stick to negatively charged grains much longer than the neutral BE predicts. H$_2$S thus only evaporates at (somewhat) higher temperatures compared to the neutral case.

Next, the electron-accepting capabilities of OCS are linked to the high structural reorganisation that must occur before it can accept an electron. Hence, the electron-accepting behaviour of OCS strongly depends on the surface morphology and the site at which OCS binds. The low BE of neutral OCS on both neutral and charged ASW means that it can easily desorb back into the gas phase, which is indeed in accordance with observations \citep{Palumbo1995,McClure2023,Palumbo1997,Sturm2023,Boogert2022}. Nevertheless, when OCS accepts an electron, its BE increases significantly, essentially locking it into the ice cluster. The OCS$^-$ anion is also much more reactive, leading to further processing into refractory organosulfur or other non-volatile compounds.

While SO$_2$ has been detected in the gas phase of cold molecular clouds, its abundance is much lower than in warm dense clouds and star-forming regions \citep{Cernicharo2011}. The high BE of SO$_2$ on charged surfaces due to the easy electron transfer to SO$_2$ provides a clear rationale for its low gas-phase abundance. Additionally, when the SO$_2^-$ anion is formed, it readily reacts with neighbouring species in the ice (such as H, O, or OH radicals) to create highly stable refractory sulfur compounds, thereby locking sulfur onto the ice. The charged surface acts in this case as a one-way valve for SO$_2$, converting it upon adsorption into highly bound species that are unable to desorb again.

Finally, it is known that sulfur in diffuse clouds is present in ionised atomic form, and it has been suggested that the S$^+$ cations freeze out onto negatively charged dust grains before the formation of a dense cloud \citep{Taillard2025,Fuente2023}. Following this freeze-out, they efficiently hydrogenate into H$_2$S, analogously to O atoms \citep{Caselli1994}. However, if the grain still holds a negative charge after the neutralisation of S$^+$, neutral S promptly becomes negatively charged, making it much more reactive than its neutral form. This opens up new pathways for S atoms; they do not just hydrogenate by reacting with incoming atomic H, they can actually easily interact with any open-shell species. The rapid charging and subsequent reaction efficiently lock the atomic sulfur from the diffuse clouds onto the solid phase, significantly accelerating the depletion of sulfur from the gas phase.

\section{Conclusion}

The aim of this work was to calculate statistically robust results for the BEs of H$_2$S, SO$_2$, S, and OCS on both neutral and charged ASW surfaces, assessing how a single negative charge might affect the BE distribution of these sulfur-bearing compounds. The BEs on neutral surfaces show a site-dependent spread, which is in accordance with previous works regarding the BE of interstellar molecules on icy dust grains. On charged surfaces, three cases were found for the behaviour of the adsorbing molecule. First, there is always a charge transfer due to the high positive electron affinities of SO$_2$ and S. Second, there is no electron transfer due to the negative electron affinity of H$_2$S. Third, an electron transfer can occur in specific cases where the structural reorganisation of OCS to accept the extra electron is favourable. It can thus be concluded that the effect of a negative charge on these grains is molecule-specific rather than a general effect. Due to this molecule-specific change in BEs, an explicit approach for each molecule is necessary in gas--grain models. Both SO$_2$ and S readily become charged when adsorbed on a charged grain, making them much more reactive to forming highly bound species on the dust grains. OCS is less likely to accept the extra electron, and it desorbs easily into the gas phase due to the low BEs on both neutral and charged surfaces. However, when OCS binds in such a way that it is favourable to overcoming the structural reorganisation barrier, it again becomes very reactive and stays locked on the grains. The BE of H$_2$S increases slightly due to the dipole interaction between H$_2$S and the loosely bound electron on the ASW surface. The increased BE makes H$_2$S stick longer to the grain than previously assumed when only neutral surfaces were taken into account. These findings support the idea that the missing sulfur is locked in icy dust grains in dense molecular clouds and subsequently in the protoplanetary discs originating from these dense clouds.

\begin{acknowledgements}
The resources and services used in this work were provided by the VSC (Flemish Supercomputer Centre), funded by the Research Foundation - Flanders (FWO) and the Flemish Government. This work was supported by the Marie Skłodowska–Curie Postdoctoral Fellowship (Grant Agreement No. 101211724).
\end{acknowledgements}

\bibliographystyle{aa}
\bibliography{References}

\begin{appendix}
\section{Benchmark}

Tables \ref{table:benchmark_H2S} – \ref{table:benchmark_SO2} show an overview of the extensive benchmark performed for the calculation of the BEs of the different studied molecules. The tables show the method used, the neutral binding energy ($\text{BE}^0$), the charged binding energy ($\text{BE}^{-1}$), and the difference between the neutral and charged binding energy ($\Delta\text{BE}$). Additionally, the signed error is shown for every binding energy value against the CCSD(T) value. For H$_2$S, the B3PW91/def2-TZVPPD/D3(BJ) was chosen since both the neutral and charged BEs are close to the CCSD(T) value. The $\Delta\text{BE}$ is also well represented by this method. For OCS, the chosen method ($\omega\text{B97M/def2-TZVPPD/D4REV}$) shows a good correspondence between the BE values for both neutral and charged systems. The error in $\Delta\text{BE}$ is large ($-22.57\%$), but this can be attributed to the fact that the BE does not change a lot when the system is charged, so small changes in the individual BE will have a large effect on this error. For S, RevPBE38/def2-TZVPPD/D3(0) performed best, showing the best correspondence with CCSD(T) for both neutral and charged BEs and a good correspondence for the $\Delta\text{BE}$ value. For SO$_2$, the same RevPBE38/def2-TZVPPD/D3(0) method was chosen since it performed best on the neutral BE and also showed good agreement with CCSD(T) for the charged BE and the $\Delta\text{BE}$-value.

\begin{table*}[hbp!]
\caption{H$_2$S benchmark.}
\label{table:benchmark_H2S}
\centering
\begin{tabular}{c c c c c c c c c}
\hline\hline
Method & Basis set & Disp & $\text{BE}^0$ (K) & $\text{Error}^0$ & $\text{BE}^{-1}$ (K) & $\text{Error}^{-1}$ & $\Delta\text{BE}$ (K) & $\text{Error}^\Delta$ \\
\hline
    CCSD(T) & aug-cc-pVTZ & x & $1305$ & x & $3187$ & x & $1881$ & x \\
    PBE0 & def2-TZVPPD & D4 & $1503$ & $15.15\%$ & $3896$ & $22.24\%$ & $2393$ & $27.17\%$ \\
    PBE0 & def2-TZVPPD & D3(BJ) & $1556$ & $19.21\%$ & $3900$ & $22.38\%$ & $2344$ & $24.58\%$ \\
    REVPBE38 & def2-TZVPPD & D3(BJ) & $1384$ & $6.97\%$ & $3446$ & $8.14\%$ & $2062$ & $9.62\%$ \\
    $\omega\text{B97X}$ & aug-cc-pVTZ & D3(BJ) & $1533$ & $17.41\%$ & $3566$ & $11.90\%$ & $2033$ & $8.08\%$ \\
    B3PW91 & def2-TZVPPD & D4 & $1140$ & $-12.70\%$ & $2991$ & $-6.15\%$ & $1851$ & $-1.60\%$ \\
    B1LYP & def2-TZVPPD & D4 & $1340$ & $2.65\%$ & $3164$ & $-0.71\%$ & $1824$ & $-3.05\%$ \\
    B3PW91 & def2-TZVPPD & D3(BJ) & $1291$ & $-1.15\%$ & $3077$ & $-3.46\%$ & $1786$ & $-5.07\%$ \\
    REVPBE0 & def2-TZVPPD & D3(BJ) & $1384$ & $6.00\%$ & $2993$ & $-6.08\%$ & $1609$ & $-14.46\%$ \\
    B3PW91 & aug-cc-pVTZ & D3(BJ) & $1300$ & $-0.45\%$ & $2637$ & $-17.25\%$ & $1338$ & $-28.91\%$ \\
\hline
\end{tabular}
\end{table*}

\begin{table*}[hbp!]
\caption{ OCS benchmark.}
\label{table:benchmark_OCS}
\centering
\begin{tabular}{c c c c c c c c c}
\hline\hline
Method & Basis set & Disp & $\text{BE}^0$ (K) & $\text{Error}^0$ & $\text{BE}^{-1}$ (K) & $\text{Error}^{-1}$ & $\Delta\text{BE}$ (K) & $\text{Error}^\Delta$ \\
\hline
    CCSD(T) & aug-cc-pVTZ & x & $1758$ & x & $1496$ & x & $-262$ & x \\
    $\omega\text{B97X}$ & def2-TZVPPD & D3(BJ) & $1777$ & $1.12\%$ & $1175$ & $-21.45\%$ & $-602$ & $130.24\%$ \\
    $\omega\text{B97X}$ & def2-TZVPPD & D4REV & $1823$ & $3.75\%$ & $1231$ & $-17.74\%$ & $-593$ & $126.67\%$ \\
    $\omega\text{B97X}$ & def2-TZVPPD & D4 & $1874$ & $6.60\%$ & $1357$ & $-9.32\%$ & $-517$ & $97.68\%$ \\
    $\omega\text{B97M}$ & def2-TZVPPD & D4 & $1771$ & $0.74\%$ & $1463$ & $-2.19\%$ & $-307$ & $17.50\%$ \\
    $\omega\text{B97M}$ & def2-TZVPPD & D3(BJ) & $1774$ & $0.94\%$ & $1474$ & $-1.46\%$ & $-300$ & $14.63\%$ \\
    B1LYP & def2-TZVPPD & D4 & $1908$ & $8.58\%$ & $1614$ & $7.86\%$ & $-295$ & $12.67\%$ \\
    B3LYP & def2-TZVPPD & D3(BJ) & $1859$ & $5.78\%$ & $1583$ & $5.79\%$ & $-276$ & $5.71\%$ \\
    B3LYP & def2-TZVPPD & D4 & $1881$ & $7.03\%$ & $1658$ & $10.81\%$ & $-223$ & $-14.59\%$ \\
    REVPBE38 & def2-TZVPPD & D3(BJ) & $1568$ & $-10.79\%$ & $1354$ & $-9.47\%$ & $-214$ & $-18.35\%$ \\
    $\omega\text{B97M}$ & def2-TZVPPD & D4REV & $1755$ & $-0.14\%$ & $1553$ & $3.79\%$ & $-203$ & $-22.57\%$ \\
    B3PW91 & def2-TZVPPD & D3(BJ) & $1436$ & $-18.29\%$ & $1282$ & $-14.34\%$ & $-155$ & $-40.89\%$ \\
    REVPBE38 & def2-TZVPPD & D4 & $1579$ & $-10.15\%$ & $1438$ & $-3.91\%$ & $-142$ & $-45.81\%$ \\
    PBE0 & def2-TZVPPD & D3(BJ) & $1801$ & $2.46\%$ & $1670$ & $11.65\%$ & $-130$ & $-50.11\%$ \\
    PBE0 & def2-TZVPPD & D4 & $1812$ & $3.11\%$ & $1718$ & $14.85\%$ & $-94$ & $-64.08\%$ \\
\hline
\end{tabular}
\end{table*}

\begin{table*}[hbp!]
\caption{ S benchmark.}
\label{table:benchmark_S}
\centering
\begin{tabular}{c c c c c c c c c}
\hline\hline
Method & Basis set & Disp & $\text{BE}^0$ (K) & $\text{Error}^0$ & $\text{BE}^{-1}$ (K) & $\text{Error}^{-1}$ & $\Delta\text{BE}$ (K) & $\text{Error}^\Delta$ \\
\hline
    CCSD(T) & aug-cc-pVTZ & x & $1740$ & x & $38060$ & x & $36320$ & x \\
    $\omega\text{B97M}$ & def2-TZVPPD & D3(BJ) & $2378$ & $36.69\%$ & $46130$ & $21.21\%$ & $43752$ & $20.46\%$ \\
    $\omega\text{B97M}$ & def2-TZVPPD & VV10 & $2461$ & $41.47\%$ & $46135$ & $21.22\%$ & $43673$ & $20.25\%$ \\
    $\omega\text{B97M}$ & def2-TZVPPD & D4REV & $2289$ & $31.55\%$ & $45932$ & $20.69\%$ & $43644$ & $20.17\%$ \\
    $\omega\text{B97M}$ & def2-TZVPPD & D4 & $2205$ & $26.73\%$ & $45808$ & $20.36\%$ & $43603$ & $20.05\%$ \\
    $\omega\text{B97X}$ & def2-TZVPPD & VV10 & $2412$ & $38.61\%$ & $43149$ & $13.37\%$ & $40737$ & $12.16\%$ \\
    $\omega\text{B97X}$ & def2-TZVPPD & D3(BJ) & $2531$ & $45.48\%$ & $43095$ & $13.23\%$ & $40564$ & $11.69\%$ \\
    $\omega\text{B97X}$ & def2-TZVPPD & D4 & $2508$ & $44.14\%$ & $42917$ & $12.76\%$ & $40409$ & $11.26\%$ \\
    $\omega\text{B97X}$ & def2-TZVPPD & D4REV & $2354$ & $35.32\%$ & $42755$ & $12.34\%$ & $40401$ & $11.24\%$ \\
    O3LYP & def2-TZVPPD & D4 & $3170$ & $82.22\%$ & $42467$ & $11.58\%$ & $39297$ & $8.20\%$ \\
    X3LYP & def2-TZVPPD & D4 & $2969$ & $70.65\%$ & $42116$ & $10.66\%$ & $39147$ & $7.78\%$ \\
    B3PW91 & def2-TZVPPD & D3BJ & $2633$ & $51.31\%$ & $41416$ & $8.82\%$ & $38783$ & $6.78\%$ \\
    B1LYP & def2-TZVPPD & D4 & $2630$ & $51.19\%$ & $41384$ & $8.74\%$ & $38754$ & $6.70\%$ \\
    B3PW91 & def2-TZVPPD & D4 & $2434$ & $39.90\%$ & $41132$ & $8.07\%$ & $38698$ & $6.55\%$ \\
    BHANDHLYP & def2-TZVPPD & D3BJ & $2324$ & $33.57\%$ & $40706$ & $6.95\%$ & $38382$ & $5.68\%$ \\
    PBE0 & def2-TZVPPD & D3BJ & $2826$ & $62.42\%$ & $41173$ & $8.18\%$ & $38347$ & $5.58\%$ \\
    BHANDHLYP & def2-TZVPPD & D4 & $2218$ & $27.47\%$ & $40547$ & $6.54\%$ & $38329$ & $5.53\%$ \\
    PBE0 & def2-TZVPPD & D4 & $2771$ & $59.28\%$ & $41083$ & $7.94\%$ & $38311$ & $5.48\%$ \\
    revPBE0 & def2-TZVPPD & D3BJ & $2391$ & $37.43\%$ & $40101$ & $5.36\%$ & $37710$ & $3.83\%$ \\
    revPBE0 & def2-TZVPPD & D4 & $2347$ & $34.87\%$ & $39866$ & $4.75\%$ & $37519$ & $3.30\%$ \\
    REVPBE38 & def2-TZVPPD & D3BJ & $2178$ & $25.20\%$ & $39686$ & $4.27\%$ & $37508$ & $3.27\%$ \\
    REVPBE38 & def2-TZVPPD & D3(0) & $1889$ & $8.57\%$ & $39390$ & $3.50\%$ & $37501$ & $3.25\%$ \\
    REVPBE38 & def2-TZVPPD & D4 & $2031$ & $16.72\%$ & $39400$ & $3.52\%$ & $37369$ & $2.89\%$ \\
\hline
\end{tabular}
\end{table*}

\begin{table*}[hbp!]
\caption{ SO$_2$ benchmark.}
\label{table:benchmark_SO2}
\centering
\begin{tabular}{c c c c c c c c c}
\hline\hline
Method & Basis set & Disp & $\text{BE}^0$ (K) & $\text{Error}^0$ & $\text{BE}^{-1}$ (K) & $\text{Error}^{-1}$ & $\Delta\text{BE}$ (K) & $\text{Error}^\Delta$ \\
\hline
    CCSD(T) & aug-cc-pVTZ & x & $3632$ & x & $29461$ & x & $25829$ & x \\
    M062X & def2-TZVPPD & D3(0) & $4904$ & $35.01\%$ & $33005$ & $12.03\%$ & $28101$ & $8.80\%$ \\
    $\omega\text{B97M}$ & def2-TZVPPD & D3(BJ) & $4399$ & $21.11\%$ & $32192$ & $9.27\%$ & $27792$ & $7.60\%$ \\
    BHANDHLYP & def2-TZVPPD & D4 & $4819$ & $32.67\%$ & $31701$ & $7.60\%$ & $26881$ & $4.07\%$ \\
    BHANDHLYP & def2-TZVPPD & D3(BJ) & $4932$ & $35.76\%$ & $31769$ & $7.83\%$ & $26837$ & $3.90\%$ \\
    $\omega\text{B97X}$ & def2-TZVPPD & VV10 & $4414$ & $21.50\%$ & $30996$ & $5.21\%$ & $26583$ & $2.92\%$ \\
    $\omega\text{B97X}$ & def2-TZVPPD & D4REV & $4365$ & $20.16\%$ & $30827$ & $4.64\%$ & $26462$ & $2.45\%$ \\
    $\omega\text{B97X}$ & def2-TZVPPD & D3(BJ) & $4492$ & $23.66\%$ & $30881$ & $4.82\%$ & $26389$ & $2.17\%$ \\
    $\omega\text{B97X}$ & def2-TZVPPD & D4 & $4573$ & $25.88\%$ & $30961$ & $5.09\%$ & $26388$ & $2.16\%$ \\
    B1LYP & def2-TZVPPD & D4 & $4724$ & $30.05\%$ & $30359$ & $3.05\%$ & $25635$ & $-0.75\%$ \\
    X3LYP & def2-TZVPPD & D4 & $4989$ & $37.36\%$ & $30455$ & $3.37\%$ & $25466$ & $-1.41\%$ \\
    REVPBE38 & def2-TZVPPD & D2 & $4403$ & $21.22\%$ & $29747$ & $0.97\%$ & $25343$ & $-1.88\%$ \\
    REVPBE38 & def2-TZVPPD & D3(0) & $4259$ & $17.24\%$ & $29530$ & $0.23\%$ & $25271$ & $-2.16\%$ \\
    REVPBE38 & def2-TZVPPD & D4 & $4380$ & $20.59\%$ & $29511$ & $0.17\%$ & $25131$ & $-2.70\%$ \\
    REVPBE38 & def2-TZVPPD & D3BJ & $4514$ & $24.28\%$ & $29643$ & $0.62\%$ & $25129$ & $-2.71\%$ \\
    B3PW91 & def2-TZVPPD & D4 & $4542$ & $25.03\%$ & $29432$ & $-0.10\%$ & $24890$ & $-3.63\%$ \\
    B3PW91 & def2-TZVPPD & D3BJ & $4725$ & $30.07\%$ & $29564$ & $0.35\%$ & $24839$ & $-3.83\%$ \\
    PBE0 & def2-TZVPPD & D4 & $5021$ & $38.23\%$ & $29589$ & $0.43\%$ & $24567$ & $-4.88\%$ \\
    PBE0 & def2-TZVPPD & D3BJ & $5042$ & $38.81\%$ & $29603$ & $0.48\%$ & $24560$ & $-4.91\%$ \\
    revPBE0 & def2-TZVPPD & D3BJ & $4404$ & $21.24\%$ & $28887$ & $-1.95\%$ & $24483$ & $-5.21\%$ \\
    revPBE0 & def2-TZVPPD & D4 & $4481$ & $23.36\%$ & $28883$ & $-1.96\%$ & $24402$ & $-5.52\%$ \\
    O3LYP & def2-TZVPPD & D4 & $5129$ & $41.19\%$ & $29164$ & $-1.01\%$ & $24036$ & $-6.94\%$ \\
\hline
\end{tabular}
\end{table*}

\clearpage
\FloatBarrier
\section{Comparison with literature}

\begin{table*}[hbp!]
\caption{Literature BEs of studied sulfur-bearing species on ASW.}
\label{table:literature}
\centering
\begin{tabular}{m{3.75em} m{3em} m{5em} m{4.5em} m{4.5em} m{4.5em} m{4.5em} m{13em}}
\hline\hline
Reference & Type & Model/ origin & $\text{BE}_{H_2S}$ (K) & $\text{BE}_{SO_2}$ (K) & $\text{BE}_{OCS}$ (K) & $\text{BE}_{S}$ (K) & Method \\
\hline
    This work & Theory & \small{Amorphous water ice cluster (30 H$_2$O)} & $2276 \pm 625$ & $3145 \pm 819$ & $1504 \pm 580$ & $1571 \pm 728$ & \small{See the method section} \\
    \hline
    Penteado et al. (2017)$^{(1)}$ & Exp./rec. & \small{Water ice} & $2290 \pm 90$ & $3010 \pm 110$ & $2325 \pm 95$ & $985 \pm 495$ & \small{Recommended values used in sensitivity analysis; H$_2$S, SO$_2$, and OCS estimated from desorption-temperature data; S from older modelling estimates} \\
    \hline
    Wakelam et al. (2017)$^{(2)}$ & Theory & \small{Water monomer} & $2900$ & $5000$ & $2100$ & $2600$ & \small{Gaussian09; M06-2X/aug-cc-pVTZ species-H$_2$O calculations; BE estimated using a fit to experimental binding energies} \\
    \hline
    Das et al. (2018)$^{(3)}$ & Theory & \small{Water monomer/ tetramer/ hexamer} & $1727$/ $2556$/ $3232$ & $3745$ & $1139$/ $1571$/ $1808$ & $1428$ & \small{Gaussian09; MP2/aug-cc-pVDZ water-cluster calculations; directly computed cluster BEs} \\
    \hline
    Ferrero et al. (2020)$^{(4)}$ & Theory & \small{Amorphous water ice slab model (60 H$_2$O)} & $2291$ - $3338$ ($\mu=2753$, $\sigma=437$) &  &  &  & \small{CRYSTAL17; DFT//HF-3c protocol: HF-3c geometries; B3LYP-D3(BJ)/A-VTZ* single-point energies; CP-BSSE; BE(0) from ZPE scaling} \\
    \hline
    Perrero et al. (2022)$^{(5)}$ & Theory & \small{Amorphous water ice slab model (60 H$_2$O)} & $1970$ - $4489$ ($\mu=3017$, $\sigma=884$) & $2105$ - $5827$ ($\mu=3254$, $\sigma=1089$) & $1286$ - $2861$ ($\mu=2083$, $\sigma=549$) & $1579$ - $2799$ ($\mu=2020$, $\sigma=366$) & \small{CRYSTAL17; DFT//HF-3c protocol: HF-3c geometries; B3LYP-D3(BJ)/A-VTZ* for closed-shell species and M06-2X/A-VTZ* for atomic S; CP-BSSE; BE(0) from ZPE scaling} \\
    \hline
    Bariosco et al. (2024)$^{(6)}$ & Theory & \small{ACO-FROST amorphous icy grain (200 H$_2$O)} & $57$ - $2406$ ($\mu=984$, $\sigma=532$) &  &  &  & \small{ORCA 5.0.3; Geometries and ZPE: ONIOM(B97-3c:GFN2-xTB); final single-point energies: ONIOM(DLPNO-CCSD(T)/aug-cc-pVTZ:GFN2-xTB); CP-BSSE} \\
    \hline
    Bulik et al. (2025)$^{(7)}$ & Theory & \small{Amorphous water ice slab model (60 H$_2$O)} & & & $1524$ - $3222$ ($\mu=2333$, $\sigma=613$) & & \small{CRYSTAL17; B3LYP-D3(BJ)/A-VTZ* geometry optimisation and energies; CP-BSSE; explicit ZPE correction} \\
    \hline
    Perrero et al. (2024)$^{(8)}$ & Theory & \small{Periodic water ice, flat / cavity regions} & & & & $1972$/ $3067$ & \small{CRYSTAL17; BHLYP-D3(BJ)/A-VTZ* periodic calculations; CP-BSSE; explicit BE(0) for flat and cavity sites} \\
    \hline
    KIDA$^{(9)}$ & Database & \small{Database desorption energy entries} & $2700$ & $3400$ & $2400$ & $2600$ & \small{Database values} \\
    \hline
    UMIST/ Rate12$^{(10)}$ & Database & \small{Database desorption energy entries} & $2743$ & $5330$ & $2888$ & $1100$ & \small{Database values} \\
\hline
\end{tabular}
\tablefoot{BEs are given in Kelvin. For multi-site ASW calculations, ranges are given with the mean $\mu$ and standard deviation $\sigma$ when available. Distinguishes recommended/experimental values, direct theoretical calculations, and database entries. ASW = amorphous solid water; BE = binding energy; BSSE = basis-set superposition error; CP = counterpoise correction; ZPE = zero-point energy; Exp./rec. = experimental or recommended value.}
\tablebib{$^{(1)}$\citet{Penteado2017}; $^{(2)}$\citet{Wakelam2017}; $^{(3)}$\citet{Das2018}; $^{(4)}$\citet{Ferrero2020}; $^{(5)}$\citet{Perrero2022a}; $^{(6)}$\citet{Bariosco2024}; $^{(7)}$\citet{Bulik2025}; $^{(8)}$\citet{Perrero2024b}; $^{(9)}$\citet{Wakelam2015}; $^{(10)}$\citet{McElroy2013}.}
\end{table*}

\end{appendix}
\end{document}